# Opportunities for DOE National Laboratory-led QuantISED Experiments

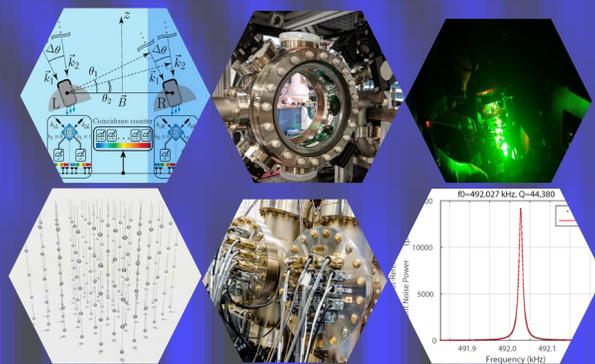

Pete Barry (ANL), Karl Berggren (MIT), A. Baha Balantekin (UW-Madison), John Bollinger (NIST), Ray Bunker (PNNL), Ilya Charaev (MIT), Jeff Chiles (NIST), Aaron Chou (FNAL), Marcel Demarteau (ORNL), Joe Formaggio (MIT), Peter Graham (Stanford), Salman Habib (ANL), David Hume (NIST), Kent Irwin (SLAC/Stanford), Mikhail Lukin (Harvard), Joseph Lykken (FNAL), Holger Mueller (UC Berkeley), Reina Maruyama (Yale), SaeWoo Nam (U. Colorado/NIST), Andrei Nomerotski (BNL), John Orrell (PNNL), Robert Plunkett (FNAL), Raphael Pooser (ORNL), John Preskill (Caltech), Surjeet Rajendran (JHU), Alex Sushkov (Boston U.), and Ronald Walsworth (U. Maryland).

# Opportunities for DOE National Laboratory-led
# QuantISED (Quantum Information Science Enabled Discovery) Experiments

Pete Barry (ANL), Karl Berggren (MIT), A. Baha Balantekin (UW-Madison), John Bollinger (NIST), Ray Bunker (PNNL), Ilya Charaev (MIT), Jeff Chiles (NIST), Aaron Chou (FNAL), Marcel Demarteau (ORNL), Joe Formaggio (MIT), Peter Graham (Stanford), *Salman Habib (ANL), David Hume (NIST), *Kent Irwin (SLAC/Stanford), Mikhail Lukin (Harvard), *Joseph Lykken (FNAL), Reina Maruyama (Yale), Holger Mueller (UC Berkeley), SaeWoo Nam (U. Colorado/NIST), *Andrei Nomerotski (BNL), John Orrell (PNNL), Robert Plunkett (FNAL), *Raphael Pooser (ORNL), John Preskill (Caltech), *Surjeet Rajendran (JHU), Alex Sushkov (Boston U.), and Ronald Walsworth (U. Maryland).

A subset of QuantISED Sensor PIs met virtually on May 26, 2020 to discuss a response to a charge by the DOE Office of High Energy Physics. Asterisks denote members of the writing group. Thanks to Monika Schleier-Smith for an invited presentation to the participants.

Photo credits: Argonne National Laboratory, A. Nomerotski et al arXiv:2012.02812, National Institute of Standards and Technology, Stanford University, University of California at Berkeley.

## Executive Summary

Quantum 2.0 advances in sensor technology offer many opportunities and new approaches for HEP experiments. The DOE HEP QuantISED program could support a portfolio of small experiments based on these advances. Such QuantISED experiments could utilize sensor technologies that exemplify Quantum 2.0 breakthroughs. They would strive to achieve new HEP science results, while possibly spinning off other domain science applications or serving as pathfinders for future HEP science targets. QuantISED experiments should be led by a DOE laboratory, in order to take advantage of laboratory technical resources, infrastructure, and expertise in the safe and efficient construction, operation, and review of experiments.

The QuantISED PIs emphasized that the quest for HEP science results under the QuantISED program is distinct in focus from the ongoing DOE HEP programs on the energy, intensity, and cosmic frontiers. There is robust evidence for the existence of particles and phenomena beyond the Standard Model, including dark matter, dark energy, quantum gravity, and new physics responsible for neutrino masses, cosmic inflation, and the cosmic preference for matter over antimatter. Where is this physics and how do we find it? The QuantISED program has opportunities to exploit new capabilities provided by quantum technology to probe these kinds of science questions in new ways and over a broader range of science parameters than can be achieved with conventional techniques.

In this document, we summarize the QuantISED sensor community discussion, including a consideration of HEP science enabled by quantum sensors, describing the distinction between Quantum 1.0 and Quantum 2.0, and discussing synergies/complementarity with the new DOE NQI centers and with research supported by other SC offices.

## Introduction

A second revolution in quantum mechanics over the last decade – sometimes referred to as the "Quantum 2.0" revolution – has led to dramatic breakthroughs in our ability to create and manipulate quantum states. These breakthroughs create new opportunities in information processing and sensing, and they can strongly impact the High Energy Physics mission. The DOE Office of High Energy Physics has been investing in this new capability through the "Quantum Information Science Enabled Discovery (QuantISED)" program.

On May 26, 2020, a group of QuantISED PIs gathered by videoconference to discuss the potential to develop and conduct experiments based on quantum sensors under the QuantISED program. This group was organized based on a charge letter from the DOE to consider "opportunities for DOE National Laboratory-led QuantISED experiments." This invitation-only workshop included participants from all of the funded sensing programs and exemplars under QuantISED.

The interdisciplinary QuantISED PIs that were gathered represent many different labs and universities, at different points in their career, and with diverse viewpoints. However, some clear common views emerged in the discussion. The focus of a QuantISED experiment is more specific than R&D in the broader QuantISED program itself, and these views can be understood to be specific to new opportunities for experiments, not as views or delimiters of the broader QuantISED program.

With that context, the participants largely coalesced around the viewpoint that QuantISED experiments would utilize sensor technologies that exemplify Quantum 2.0 technology breakthroughs. They can strive to achieve new HEP science results, while possibly spinning off other domain science applications or serving as pathfinders for future HEP science targets. The group concurs that QuantISED experiments should be led by a DOE laboratory, in order to take advantage of lab technical resources, infrastructure, and expertise in the safe and efficient construction, operation, and review of experiments. The QuantISED PIs emphasized that the quest for HEP science results under the QuantISED program is distinct in focus from the ongoing DOE HEP programs on the energy, intensity, and cosmic frontiers. The QuantISED program can exploit new capabilities provided by quantum sensor technology to probe HEP science in new ways and over a broader range of science parameters. For instance, a compelling QuantISED experiment might take advantage of quantum-enhanced sensitivity to search for a broader class of dark-matter candidates than is probed by the cosmic, energy, and intensity frontiers, while also being a steppingstone towards even more sensitive laboratory experiments to search for direct interaction with certain models of dark energy.

**HEP science and quantum sensors**

The Standard Model of particle physics has withstood every direct experimental test. Yet, there is compelling evidence that the theory is incomplete. It fails to account for known observational facts about the universe such as the existence of dark matter, dark energy, the matter-anti matter asymmetry and the physics responsible for neutrino masses. It does not describe quantum gravity or the origin of the universe. How can we find this new physics? The "dark" or weakly coupled nature of many of these phenomena are suggestive that this physics can be accessed by experiments that offer unprecedented sensitivity. This can be accomplished by making precision measurements or looking for phenomena that are rare because they involve new or highly suppressed interactions. Such experiments are ultimately limited by the sensitivity of the devices used to detect these interactions. The advent of quantum sensors offers many opportunities and new approaches to broaden and advance this sensitivity frontier. This approach is complementary to other powerful experimental tools that exist to search for new physics that may manifest at high energies with Standard Model particles.

Quantum technologies enable a diverse probe of HEP science targets, complementary to other agency missions. For example, emerging quantum techniques can enable searches for a broader class of dark matter candidates and interactions than more traditional approaches. This includes, among other examples, candidates such as axion-like-particles, relaxions and moduli that induce spin precession or affect the energy levels of atoms and molecules. Further, these techniques can search for dark matter masses ranging from 100 nHz to nearly 1 THz, distinct from other probes of dark matter. In fact, quantum sensors have already made game-changing contributions to the search for dark matter, enabling ongoing and planned searches for a significantly larger range of dark sector targets than previously possible.

Quantum techniques that can coherently control and manipulate photons, spins, atoms and molecules can be used to advance multiple HEP goals. This includes the use of : (1) electromagnetic sensors to look for dark matter and new fundamental interactions, for example, through high Q electromagnetic devices,  (2) spin sensors to search for the precession induced by dark matter, dark energy and terrestrial neutrino sources, (3) atomic sensors and clocks to search for dark matter and new fundamental interactions (4) molecules and quantum materials to probe sources of CP violation in the universe and probe dark matter (5) photon teleportation techniques to increase baselines and hence the sensitivity of optical interferometers or other quantum sensing networks by orders of magnitude. This list is by no means exhaustive. The current state of quantum sensing already allows for immediate exploration of new parameter space in the search for dark matter, new fundamental interactions and CP violation.

Importantly, quantum-enabled experiments that aim to achieve a science result for one HEP science target can simultaneously serve as a pathfinder for another HEP science target. For example, an experiment that aims to sense the spin precession induced by dark matter could also demonstrate and mature the technology needed for the even more ambitious goal of direct detection of dark energy in the laboratory.  Ultimately, these technologies could also lead to new ways to detect a current of neutrinos, enabling better calibration of current neutrino sources and the detection of very low energy neutrinos, a task that is otherwise difficult to accomplish. Similarly, new approaches to detecting dark matter could also create a path towards probing cosmic inflation by searching for relic particles produced during inflation. Experiments that look for new fundamental interactions or violations of fundamental symmetries through the effects of such physics on the energy levels of atoms and molecules can also serve as pathfinders for similar, time-varying effects caused by dark matter.

## Quantum 2.0 vs. Quantum 1.0

The QuantISED sensor community broadly feels that new science experiments conducted under the QuantISED program would best take advantage of the new breakthroughs that are part of Quantum 2.0. But we must be careful to define what we mean by Quantum 2.0, to avoid missing important new quantum techniques that might not fit a too-narrow definition. There was some concern expressed in the workshop that emphasizing the use of Quantum 2.0 sensors without clarifying this distinction might lead to greater confusion.

The authors of this QuantISED study feel that three definitions, together, provide good coverage of relevant techniques:
- (Type 1) Experiments that evade one of several clearly defined Standard Quantum Limits (SQL) of measurement by, for instance, using phase-sensitive techniques (such as squeezing or backaction evasion), or by taking advantage of entanglement in measurement. For example, dc SQUIDs clearly are limited by the SQL since they are phase-insensitive amplifiers, but other Josephson-junction-based devices like qubits, parametric amplifiers or upconverters can be used to evade the SQL.
- (Type 2) Experiments that make use of the ability to prepare, control, transmit and measure coherent quantum states provided by recent breakthroughs in the QIS community (relevant systems include Josephson qubits/devices, clocks, atom interferometers, NMR systems, color centers, optical interferometers and other photonics systems, etc…). Relevant techniques of this type were unknown or only theoretical 30 years ago.
- (Type 3) Some Quantum 2.0 techniques are defined by their use. For example, bolometric photon detectors may not be Quantum 2.0 on their own, but are key enablers of Quantum 2.0 experiments such as detecting entangled photon pairs for Bell State Measurements or Quantum Key Distribution.

## Synergies with the new DOE NQI centers and with research supported by other SC offices

The community believes that the newly established DOE NQI centers will be a primary method of cross-cutting collaboration benefiting fundamental HEP science; indeed, some of the centers have explicit HEP science goals in their scope, and all five centers have quantum sensor goals relevant to HEP applications. Future HEP work supported by QuantISED will be different from and benefit from work ongoing in NQI centers.

The QuantISED PI community saw multiple ways to link together with research supported by other SC offices. The meeting attendees noted that recent Quantum 2.0 experiments for DOE BES have some commonality with tabletop HEP experiments. Therefore, it is possible that BES experiments that seek to go below the SQL, for example, can benefit from HEP experimental techniques that currently use entanglement and quantum noise reduction. Likewise, certain demonstrations of going below the SQL in BES may be applicable to HEP demonstrator or tabletop experiments. There is also overlap with possible quantum experiments in NP to probe neutrinos and fundamental symmetries. We can also leverage the algorithmic work being done in ASCR. The focus so far has been on developing algorithms for quantum computers, with application to the domain sciences, but little investment has been made into algorithms for quantum sensing that would benefit HEP experiments. A cadre of QuantISED PIs have begun collaborating with quantum computer experts to produce these algorithms in recent years, and this should continue, taking advantage of the NQI centers as a natural venue.

Finally, this group believes that quantum networking will play a role in some future HEP experiments. HEP and ASCR have complementary interests in quantum networks, as described in the DOE quantum internet blueprint report. For example, networking quantum computers is an ASCR interest, while networking quantum devices for the purposes of correlating signals is an HEP interest. Both long range networks of interest to ASCR and quantum sensor arrays of interest to HEP will benefit from ongoing research in quantum information transduction (RF to optical and even audio, for example). Finding the synergies between these methods will lead to a natural partnership between programs at the SC level.

Strong connections between the QuantISED research community and the larger quantum science research ecosystem is also important for quantum workforce development, including the key challenge of attracting and retaining people from diverse backgrounds in a field currently notable for its lack of diversity.